\documentclass
[prl,lengthcheck,10pt,tightenlines,twocolumn,superscriptaddress,footinbib]{revtex4}%
\usepackage{amsfonts}
\usepackage{amsmath}
\usepackage{amssymb}
\usepackage{graphicx}%
\providecommand{\U}[1]{\protect\rule{.1in}{.1in}}

\begin{document}
\title{Basin Entropy in Boolean Network Ensembles}
\author{Peter Krawitz}
\affiliation{Institute for Systems Biology, Seattle, WA 98103, USA }
\affiliation{Fakult\"{a}t f\"{u}r Physik, Ludwig Maximilians Universit\"{a}t, 80799
M\"{u}nchen, Germany}
\author{Ilya Shmulevich}
\affiliation{Institute for Systems Biology, Seattle, WA 98103, USA }

\begin{abstract}
The information processing capacity of a complex dynamical system is reflected
in the partitioning of its state space into disjoint basins of attraction,
with state trajectories in each basin flowing towards their corresponding
attractor. We introduce a novel network parameter, the basin entropy, as a
measure of the complexity of information that such a system is capable of
storing. By studying ensembles of random Boolean networks, we find that the
basin entropy scales with system size only in critical regimes, suggesting
that the informationally optimal partition of the state space is achieved when
the system is operating at the critical boundary between the ordered and
disordered phases.

\end{abstract}
\maketitle

\paragraph{Introduction}

Complex dynamical systems generically possess state spaces that are
partitioned into disjoint basins of attraction. Within each basin, state
trajectories flow to a single attractor, which, in general, may be a
steady-state fixed point, an oscillation -- including limit cycles, or a
strange (chaotic) attractor. Such classification of states via attractors and
their basins of attraction constitutes the information processing ability of a
complex system and represents a type of associative memory: all states in a
basin are in the same class in that they are all associated with the same
attractor. The complexity of information that such a system is capable of
processing depends on the manner in which its state space is partitioned,
which in turn is largely determined by the dynamical regime of the system. By
using a novel network parameter, the basin entropy, and the model class of
random Boolean networks, for which there exist well-defined and extensively
studied notions of ordered, critical, and chaotic dynamics \cite{DerridaEPS86}%
,\cite{FlyvbjergJPA88b},\cite{LuquePA00},\cite{Aldana03}, we show here that
the informationally optimal partition of the state space is achieved when the
system is operating at the critical boundary between the ordered and
disordered regimes.

Random Boolean networks (RBNs) are commonly studied as generic models of
dynamically interacting entities, such as gene regulatory networks. A RBN,
originally introduced by Kauffman \cite{KauffmanJTB69} (also called Kauffman
networks), consists of $n$ nodes, each of which can have two possible values
($0$ and $1$). Each node receives input from $k$ randomly chosen nodes that
determine its value at the next time step via a randomly chosen Boolean
function assigned to that node. The output of the function is chosen to be $1$
with probability $p$, known as the bias \cite{BastollaPD98}. A state of the
network is the collective activity of the nodes. All nodes are updated
synchronously and the network transitions from one state to another, thus
tracing out a state trajectory that eventually flows into a series of
periodically recurring states called an attractor. The transient states that
flow into an attractor constitute the basin of attraction for that attractor.

In the limit of large $n$, RBNs exhibit a phase transition between a
dynamically ordered and chaotic regime. Depending on the parameters $k$ and
$p,$ small perturbations die out over time in the ordered regime and increase
exponentially in the chaotic regime (see e.g., \cite{Aldana03}). Networks that
operate at the boundary between the ordered and the chaotic phase have been of
particular interest as models for gene regulatory networks, as they exhibit
complex dynamics combined with stability under perturbations \cite{Kauffman93}%
,\cite{RamoJTB06},\cite{ShmulevichPNAS05},\cite{SerraJTB04}. The average
length of attractors have been extensively studied numerically and
analytically for a wide range of Kauffman networks. In the chaotic phase, the
average length of attractors increases exponentially with system size
\cite{BastollaPD98}. In the highly chaotic case, where the state space of a
RBN can be approximated by a random map, the expected number of attractors
increases linearily with system size \cite{Kruskal54}. In the ordered phase,
where the fraction of nodes that freeze to a constant value approaches one,
the average number and length of attractors is bounded \cite{FlyvbjergJPA88b}%
,\cite{SocolarPRL03}. In contrast to former assumptions, it has been recently
shown that the average number and length of attractors increases
superpolynomially in critical Kauffman networks \cite{SamuelssonPRL03}.

In this Letter, we introduce a new network parameter, the basin entropy $h$ of
a Boolean network, and show that this quantity increases with system size in
critical ensembles, whereas it reaches a fixed value in the ordered
\textit{and} in the highly chaotic phase. The partition of the state space
into basins of attraction induces a probability mass function over the state
space, with the weight of each basin defined by its size relative to the other
basins. The basin entropy (hereafter, simply entropy) of a network is then
calculated from this probability distribution. The partition that a given
network imposes on its state space is not unique and thus, neither is its
entropy. Certain network instances in the ordered phase may have the same
entropy as some networks in the chaotic phase, and thus it becomes necessary
to study the average entropies of network ensembles.

In the informational sense, the entropy of a Boolean network is a measure of
the uncertainty about its dynamical behavior. In a network with higher
entropy, more information is needed to describe its future behavior and its
effective complexity is higher \cite{GellMannC96}. Thus, if the network is
viewed as a classifier, then the entropy is a measure for the complexity of
this classification process. Herein, we show that an increase in such
classification complexity as a function of system size occurs only when the
system is poised at the boundary between an ordered and disordered phase.

A Boolean Network $B$ $=(\mathbf{N,F})$ is defined by the set $\mathbf{N}$ of
its nodes, $\mathbf{N}=\left\{  1,...,n\right\}  $, and the set $\mathbf{F}$
of their corresponding Boolean updating functions $\mathbf{F}=\left\{
f_{1},...,f_{n}\right\}  $, with $f_{i}$ $:$ $\left\{  0,1\right\}
^{k}\rightarrow\left\{  0,1\right\}  $. The value $x_{i}$ of node $i$ at time
$t+1$ is determined by the values of its $k$ controlling elements as
$x_{i}(t+1)=f_{i}(x_{i(1)}(t),...,x_{i(k)}(t))$.

The sensitivity $s$ of a network, defined as the average sensitivity of the
Boolean functions used in the network \cite{ShmulevichPRL04}, is an order
parameter that specifies the average number of nodes that are affected by a
perturbation of a random node. Thus, an average sensitivity of $s=1$ indicates
that a perturbation of a random node is on average propagated to one other
node. This defines the point of phase transition between the ordered regime
($s<1$), where perturbations die out over time, and the chaotic regime
($s>1$), where even small perturbations increase over time \cite{Aldana03}. In
classical Kauffman networks, the relationship between the network parameters
$k$ and $p$ and the average sensitivity $s$ is \cite{ShmulevichPRL04}:%

\begin{equation}
s=2kp(1-p). \label{av_sens}%
\end{equation}
The logarithm of this parameter may also be interpreted as the Lyapunov
exponent, $\lambda=\log s$ \cite{LuquePA00}. In the following, we study
ensembles $\mathbf{B(}n,k,p)$ of Boolean networks with $n$ nodes parameterized
by $k$ and $p$. For $p=\frac{1}{2}$ all networks of the ensemble have equal
probability. In the general case, where certain Boolean functions might be
chosen with different probabilities, we indicate by $\upsilon_{i}$ the
probability of a certain network instance $i$ in the ensemble. We will also
find it convenient to refer to the ensemble of networks $\mathbf{B(}n,s)$ with
average sensitivity $s$.

\paragraph{Average basin entropy}

\begin{figure}
[ptb]
\begin{center}
\includegraphics[
trim=0.000000in 0.000000in 0.003814in 0.000000in,
height=2.7657in,
width=3.371in
]%
{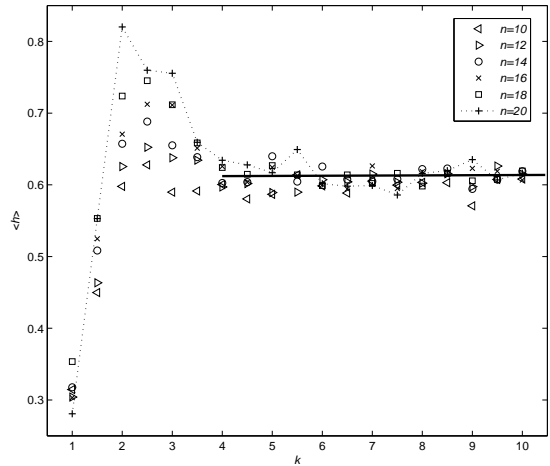}%
\caption{Average entropy in ensembles $\mathbf{B(}n,k,p=1/2)$ with
$n=10,..,20$ and $k=1,...,10$. The average entropy grows with $n$ for the
critical ensembles $\mathbf{B(}n,k=2,p=1/2)$ and approaches a finite value in
the chaotic regime independent of $n.$}%
\label{figure1}%
\end{center}
\end{figure}

Any Boolean network $B\in\mathbf{B}(n,s)$ partitions its state space into
attractors and corresponding basin states. The weight $w_{\rho}$ of an
attractor $\rho$ is the length of its attractor plus the number of basin
states draining into that attractor, normalized by the size of the state space
($2^{n}$), so that $\sum_{\rho}w_{\rho}=1$. The basin entropy $h$ of a network
$B$ is defined as:%

\begin{equation}
h(B)=-\sum_{\rho}w_{\rho}\ln w_{\rho}.
\end{equation}
The average entropy $\left\langle h\right\rangle $ of an ensemble
$\mathbf{B(}n,s)$ of networks is defined as:%

\begin{equation}
\left\langle h\right\rangle \left[  \mathbf{B(}n,s)\right]  =-\sum
_{i\in\mathbf{B(}n,s)}\upsilon_{i}\sum_{\rho_{i}}w_{\rho_{i}}\ln w_{\rho_{i}},
\end{equation}
where $\upsilon_{i}$ is the probability of a network instance in the ensemble.
To determine the state space partition of a network exactly, one has to link
each state to its attractor. We performed an exhaustive computation of the
state space partition to estimate the average entropy of the ensembles
$\mathbf{B(}n,k,p=1/2)$, with $n=10,\ldots,20$ and $k=1,\ldots,10$. As shown
in Fig. (\ref{figure1}), the average entropies of the critical ensembles $k=2$
grow with $n$ whereas the average entropies of the chaotic ensembles $k>2$
approach a finite value of approximately $0.61$, independent of $n$. Before
considering the scaling behavior of the average entropy in critical network
ensembles $\mathbf{B(}n,s=1)$, we will discuss the average entropy in the
chaotic regime in the limit of large $n$.%

When $k$ reaches $n,$ the ensemble $\mathbf{B(}n,k=n,p=1/2)$ can be identified
with the random map model, which is a simple disordered system with
deterministic dynamics. For each state in the state space, another state (not
necessarily a different one) is randomly chosen as its successor. Thus, a
random map can be regarded as an unbiased Boolean network where each node
depends on all $n$ variables. The average sensitivity of the random map is
thus $s=n/2$ and its average entropy can be expressed as:%

\begin{equation}
\left\langle h\right\rangle \left[  \mathbf{B(}n,s=n/2)\right]  =-\sum
_{j=1}^{2^{n}}g_{j}w_{j}\ln w_{j}%
\end{equation}
where the sum is taken over all possible weights $w_{j}\in\left\{
1/2^{n},2/2^{n}-1...,1\right\}  $, multiplied by the normalized frequency
$g_{j}$ of their occurrence in the ensemble $\mathbf{B(}n,s=n/2)$.

As $n\rightarrow\infty$, we may replace the sum by an integral, where $g(w)$
$dw$ indicates the average number of attractors with\ a weight between $w$ and
$w+dw$:%

\begin{equation}
\lim_{n\rightarrow\infty}\left\langle h\right\rangle \left[  \mathbf{B(}%
n,s=n/2)\right]  =-\int_{0}^{1}g(w)w\log(w)dw. \label{eq_entropy_RM}%
\end{equation}
Derrida derived the following analytical expression for the distribution of
$g$ \cite{DerridaJP88}: $g(w)=\frac{1}{2}w^{-1}(1-w)^{-\frac{1}{2}}$.
Substituting this expression into Eq. (\ref{eq_entropy_RM}) yields
$\lim_{n\rightarrow\infty}\left\langle h\right\rangle \left[  \mathbf{B(}%
n,s=n/2)\right]  =2\left(  1-\ln2\right)  =0.613\,71,$ which is the average
entropy of the random map in the limit of large $n$. It is remarkable that
ensembles of small network sizes already approach this value soon after
entering the chaotic regime (see horizontal black line in Fig. \ref{figure1}).

Before discussing the scaling behavior in critical ensembles, let us briefly
recapitulate the concept of irrelevant and relevant nodes in a network
\cite{FlyvbjergJPA88b},\cite{BastollaPD98} and their meaning in the context of
entropy. A node that is updated by a constant Boolean function is a frozen
node. All nodes, that eventually take the same constant value on every
attractor are called clamped nodes \cite{SocolarPRL03}. These nodes build the
frozen core of the network. A relevant node is one that eventually influences
its own state. The number and the length of the attractors in a network can be
determined from the set of relevant nodes.

Nonfrozen nodes that do not influence their own state are irrelevant for the
attractor dynamics and are therefore called non-frozen irrelevant nodes. As
soon as the network reaches an attractor, the future behavior of such a
nonfrozen irrelevant node can be determined as a function of only the relevant
nodes. When we add a nonfrozen irrelevant node to a network, the number of
states flowing into each of the attractors doubles. Thus, the attractor
weights do not change by adding or removing nonfrozen irrelevant nodes.
Further, the addition of a frozen node does not, on average, change the
attractor weight distribution. Thus, the average entropy of an ensemble only
depends on the organization of its relevant nodes. A set including all
relevant nodes can be obtained by iteratively removing nodes that freeze and
become part of the frozen core and non-frozen nodes that only influence
irrelevant nodes. The remaining set is also referred to as the computational
core of a network and only exists in the critical and chaotic regime with high
probability \cite{CorrealeJSM06},\cite{CorrealePRL06}. The organisation of the
relevant components is crucial for the understanding of the entropy $h$, as
the entropy of the entire network can be calculated from the entropy of the
independent non-connected relevant components $j$:%

\begin{equation}
h=\sum_{j}h_{j}%
\end{equation}

For networks in the ordered regime, the number of relevant nodes approaches a
finite limit for large system sizes \cite{MihaljevPRE06} and the proportion of
frozen nodes in the network approaches one \cite{FlyvbjergJPA88b}. Therefore,
the mean number of attractors, and thus the average entropy, is bounded for
large systems. These results indicate that increasing the system size does
not, on average, increase the entropy of networks in the ordered or chaotic
regimes, implying that complexity of classification cannot be increased in
these regimes.

\paragraph{Critical ensembles}

We first treat the special case of a critical network with connectivity $k=1$.
In these networks, all nodes are updated by either the Boolean `copy' or
`invert' function. The topology of such networks consists of loops and trees
rooted in loops. Only nodes on loops are relevant. Frozen nodes do not exist
in these networks. Consequently, attractors of the same length have the same
weight. The entropy of a critical $k=1$ network is the sum of the entropies of
its loops. In the limit of large $n,$ the number of relevant nodes scales as
$n_{r}\sim\left(  \frac{\pi n}{2}\right)  ^{\frac{1}{2}}$ (a critical $k=1$
network is a random mapping digraph and the number of relevant nodes is
equivalent to the number of vertices on cycles; see, e.g. \cite{HarrisAMS60}
and \cite{Bollobas}, chapter 14). The probabilities of having $n_{l}$ loops of
size $l$ are independent and Poisson distributed with mean $\lambda=l^{-1}$
\cite{DrosselPRL05}. If we approximate the entropy of a loop of size $l$ by
$h_{l}\approx l\ln2$ and take the sum over the expected number of such loops
for $l\leq n_{r}$, we get the following scaling behavior for the entropy in
critical $k=1$ networks \footnote[1]{In a loop, the length $l$ refers to the
number of relevant nodes. A loop with an even/odd number of `invert' functions
is called an even/odd loop. In an even loop of prime length $l$, there are
$\frac{2^{l}-2}{l}$ attractors of length $l$ and $\frac{2^{l}-1}{2l}$
attractors of length $2l$ in the odd case. If $l$ is not prime, then there are
additional attractors of shorter length (for a more detailed description see a
text book of Combinatorics about de Bruijn Sequences and Lyndon Words or see
\cite{deBruijn46}). We approximated the entropy of a loop by considering only
the `longest' attractors and considering even and odd loops as equally likely:
$h(l)\approx\ln$($\frac{2^{l}}{l}-2)+\frac{1}{2}\ln$($\frac{2^{l}}{2l}-1)$
$\approx l\ln2-\ln\sqrt{2}l.$}:%

\begin{equation}
\left\langle h\right\rangle \sim\sum_{l=1}^{n_{r}}\frac{h_{l}}{l}=\ln
2\cdot\left(  \frac{\pi n}{2}\right)  ^{\frac{1}{2}}.
\end{equation}
%

\begin{figure}
[ptb]
\begin{center}
\includegraphics[
height=2.7657in,
width=3.371in
]%
{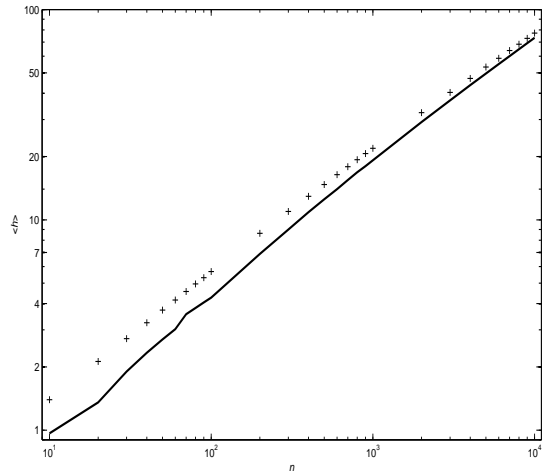}%
\caption{Average entropy $\left\langle h\right\rangle $ in critical $k=1$
networks. The symbols `+' indicate the results from simulations. $\left\langle
h\right\rangle $ approaches $\sqrt{\frac{\pi n}{2}}\ln2-\sum_{i=0}^{n_{r}%
}\frac{\ln\sqrt{2i}}{i}$ from above.}%
\label{figure2}%
\end{center}
\end{figure}

\begin{figure}
[ptb]
\begin{center}
\includegraphics[
height=2.7657in,
width=3.371in
]%
{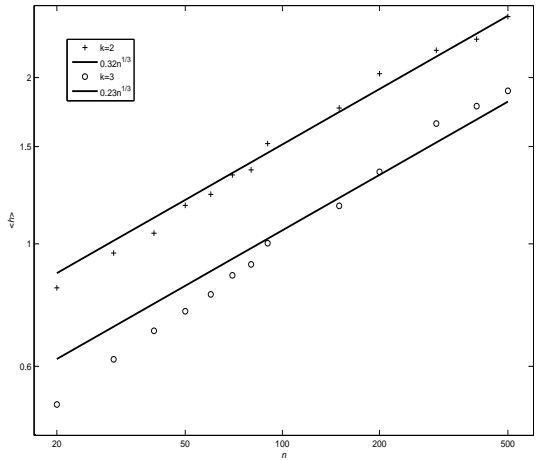}%
\caption{Average entropy of critical $k=2$ and $k=3$ ensembles, estimated from
random sampling of the state space. The data were fitted to the scaling
behavior of relevant nodes.}%
\label{figure3}%
\end{center}
\end{figure}

In networks with a connectivity higher than one, relevant nodes may depend on
more than just one relevant node. In addition to simple loops, such networks
may contain complex relevant components with crosslinks. The scaling behavior
of relevant nodes in a general class of critical random Boolean networks and
structural properties of the complex relevant components were recently
discussed in an extraordinary series of publications by Drossel, Kaufman and
Mihaljev \cite{KaufmanPRE05},\cite{KaufmanNJP06},\cite{MihaljevPRE06}. The key
result of their work is that the number of relevant nodes $n_{r}$ scales as
$n_{r}\sim n^{\frac{1}{3}}$ with the system size in all critical ensembles
with $k>1$. Furthermore, the proportion of relevant nodes that depend on more
than one relevant input approaches zero with growing $n$. From Theorem 1.3 by
Cooper and Frieze in \cite{CooperCPC04} we find that the expected number of
nodes on simple loops increases if the probability of relevant nodes depending
on one relevant input increases. As we already discussed the scaling behavior
of the average entropy of simple loops, it follows that the average entropy in
all critical ensembles increases with system size. Fig. \ref{figure3} shows
the scaling behavior of the estimated average entropy of the critical
ensembles $k=2$ and $k=3$ for $n=20,\ldots,500$ (the weight distribution of
every network was estimated by linking over 2000 random states to their
attractors and the entropy was then averaged over more than 2000 random
networks from each ensemble). For both ensembles, we fitted the data to an
expected number of $n_{r}\sim n^{\frac{1}{3}}$ relevant nodes. In both cases,
the average entropy grows with the expected number of relevant nodes. In most
cases a network will only contain one giant complex component. Relevant nodes
in this component that depend on more than one input will influence the
entropy of this component and may also explain the different average entropy
in different critical ensembles.%

\paragraph{Conclusions}

We have defined the average basin entropy of an ensemble of Boolean networks
and examined this parameter for networks in the ordered, critical and chaotic
regimes. We showed that the average entropy increases with system size in
several critical network ensembles, but not in the ordered or chaotic regimes,
suggesting that this is a general property of critical networks under
synchronous updating schemes.

Our findings are particularly relevant to the study of biological networks, as
the information processing in living systems is based on massively parallel
dynamics in complex molecular networks that underly ontogeny, immune responses
and cognitive behaviour. Kauffman hypothesized that attractors of genetic
regulatory networks can be viewed as cell types \cite{Kauffman93}. If so, then
cell types, like classes in parallel processing networks, drain basins of
attraction. Recent experimental evidence supports this hypothesis
\cite{HuangPRL05}. 

The cell is also able to respond to changes in its environment by changing its
own behaviour. Examples of such decisions include initiation of the cell
division (cell cycle), execution of specific functions (differentiation),
programmed cell death (apoptosis), and other cellular functional states. Such
decision-making motivates the view that a living system "classifies" its
environment according to its steady states. When viewed in terms of
information processing, a cell "reads" the information in its environment,
propagates that information through its intracellular networks of interacting
biomolecules and responds by exhibiting a corresponding steady-state.The
information in the steady-state reflects the information in the environment.
If maximal classification complexity confers a selective advantage, our
results lend also support to the long-standing hypothesis that living systems
are critical. 

The classical Kauffman network ensembles studied herein have characteristics
that do not reflect certain aspects of biological networks. First, the
synchronous updating scheme does not reflect the fact that elements in
biomolecular networks exhibit different time scales. Second, the organization
of biological networks is highly modular, containing many loosely coupled
modules of similar sizes. In contrast, most relevant nodes in random Boolean
networks are, with high probability, organized into one giant component. Since
the total entropy is the sum of the entropies of the components, a growing
number of such regulatory modules will certainly increase entropy, independent
of the updating scheme. Thus, understanding the entropy of complex relevant
components under different updating rules is an important topic for future studies.

\paragraph{Acknolwedgment}

This work was supported by NIH/NIGMS GM070600, GM072855, P50-GM076547 and by
the Max Weber-Programm Bayern.


\begin{thebibliography}{99}                                                                                               %


\bibitem {DerridaEPS86}B.Derrida, Y. Pomeau. \textit{Europhys. Lett.
}\textbf{1, }45 (1986)

\bibitem {FlyvbjergJPA88b}H. Flyvbjerg, \textit{J. Phys. A. }\textbf{21,
}L955-L960, 1988

\bibitem {LuquePA00}B. Luque and R. V. Sol\'{e}. \textit{Physica A}
\textbf{284}, 33-45 (2000).

\bibitem {Aldana03}M. Aldana, S. Coppersmith, and L. P. Kadanoff. in
\textit{Perspectives and Problems in Nonlinear Science}, eds. Kaplan, E.,
Marsden, J. E. \& Sreenivasan, K. R. (Springer, New York), 23-89. (2002)

\bibitem {KauffmanJTB69}S. A. Kauffman. \textit{J. Theor. Biol.}\textbf{\ 22},
437-467 (1969)

\bibitem {BastollaPD98}U. Bastolla and G. Parisi. \textit{Physica
}\textbf{115D}, 203 (1998)

\bibitem {Kauffman93}S. A. Kauffman. \textit{The origins of order:
Self-organization and selection in evolution}, Oxford University Press, New
York. (1993).

\bibitem {RamoJTB06}P. R\"{a}m\"{o}, J. Kesseli, O. Yli-Harja. \textit{J.
Theor. Biol. }\textbf{206}, 164-170 (2006)

\bibitem {ShmulevichPNAS05}I. Shmulevich, S. Kauffman, M. Aldana.
\textit{Proc. Natl. Acad. Sci. U.S.A. }\textbf{102}, 13439-13444 (2005)

\bibitem {SerraJTB04}R, Serra, M. Villani, A. Semeria. \textit{J. Theo. Biol.
}\textbf{227, }149-157 (2004)

\bibitem {DrosselPRL05}B. Drossel, T. Mihaljev, F. Greil. \textit{Phys. Rev.
Lett. }\textbf{94,} 088701 (2005)

\bibitem {KaufmanPRE05}V. Kaufman, T. Mihaljev, B. Drossel. \textit{Phys. Rev.
E. }\textbf{72, }046124 (2005)

\bibitem {SocolarPRL03}J. Socolar, S. Kauffman, \textit{Phys. Rev. Lett.
}\textbf{90, }068702 (2003)

\bibitem {GellMannC96}M. Gell-Mann, S. Lloyd. \textit{Complexity, }\textbf{1,
}44-52 (1996)

\bibitem {ShmulevichPRL04}I. Shmulevich, S. Kauffman. \textit{Phys. Rev. Lett.
}\textbf{93, }048701 (2004)

\bibitem {DerridaJP88}B. Derrida. \textit{J. Phys. }\textbf{48, }971-978 (1988)

\bibitem {KaufmanEPJB05}V. Kaufman, B. Drossel. \textit{Euro. Phys. J. B.
}\textbf{43, }115-124 (2005)

\bibitem {HarrisAMS60}B. Harris. \textit{A. Math. Stat. }\textbf{31,
}1045-1062 (1960)

\bibitem {Bollobas}B. Bollobas. \textit{Random Graphs, }Cambridge studies in
advanced mathematics, Cambridge (2001)

\bibitem {PaulPRE06}U. Paul, V. Kaufman, B. Drossel. \textit{Phys. Rev. E.
}\textbf{73, }026118 (2006)

\bibitem {Jacob61}F. Jacob, J. Monod. \textit{On the regulation of gene
activity, }Cold Spring Harbor Symposia on Quantitative Biology (1961)

\bibitem {HuangPRL05}S. Huang, \ G. Eichler, Y. Bar-Yam, d. Ingber.
\textit{Phys. Rev. Lett. }\textbf{94, }128701 (2005)

\bibitem {CooperCPC04}C. Cooper, A. Frieze. \textit{Comb. Prob. Comp.
}\textbf{13, }319-337 (2004)

\bibitem {deBruijn46}N. de Bruijn. \textit{Koninklijke Nederandse Akademie v.
Wetenschappen, }\textbf{49, }578-764 (1946)

\bibitem {KaufmanNJP06}V. Kaufman, B. Drossel. \textit{New. J. Phys.
}\textbf{8, }1 (2006)

\bibitem {MihaljevPRE06}T. Mihaljev, B. Drossel. \textit{Phys. Rev. E.
}\textbf{74, }046101 (2006)

\bibitem {Kruskal54}M. Kruskal.\textit{ A. Math. Monthly}, \textbf{61},
392-397 (1954)

\bibitem {SamuelssonPRL03}B. Samuelsson, C. Troein, \textit{Phys. Rev. Let}.
\textbf{90}, 098701 (2003)

\bibitem {GreilPRL05}F. Greil, B. Drossel, \textit{Phys. Rev. Let}.
\textbf{95}, 048701 (2005)

\bibitem {CorrealePRL06}L. Correale, M. Leone, A. Pagnani, M. Weigt, R.
Zecchina, \textit{Phys. Rev. Let}. \textbf{96}, 018101 (2006)

\bibitem {CorrealeJSM06}L. Correale, M. Leone, A. Pagnani, M. Weigt, R.
Zecchina, \textit{J. Stat. Mech}, \textbf{P03002} (2006)
\end{thebibliography}
\end{document}